\documentclass{INTERSPEECH2023}

\usepackage{multirow}
\usepackage{balance}

\usepackage{hyperref}
\usepackage{xcolor,colortbl}

\definecolor{baseline_color}{gray}{0.92}

\newcommand{\alovc}{ALO-VC}
\newcommand{\resemblyzerVC}{ALO-VC-R}
\newcommand{\ecapatdnnVC}{ALO-VC-E}

\interspeechcameraready

\title{ALO-VC: Any-to-any Low-latency One-shot Voice Conversion}\name{Bohan Wang$^1$, Damien Ronssin$^2$, Milos Cernak$^2$}
\address{
  $^1$École Polytechnique Fédérale de Lausanne, Lausanne, Switzerland\\
  $^2$Logitech Europe S.A., Lausanne, Switzerland}
\email{bohan.wang@epfl.ch, damien.ronssin@gmail.com, milos.cernak@ieee.org}

\begin{document}

\maketitle
 
\begin{abstract}

This paper presents \alovc{}, a non-parallel low-latency one-shot phonetic posteriorgrams (PPGs) based voice conversion method. \alovc{} enables any-to-any voice conversion using only one utterance from the target speaker, with only 47.5 ms future look-ahead. The proposed hybrid signal processing and machine learning pipeline combines a pre-trained speaker encoder, a pitch predictor to predict the converted speech's prosody, and positional encoding to convey the phoneme's location information. We introduce two system versions: \resemblyzerVC{}, which uses a pre-trained d-vector speaker encoder, and \ecapatdnnVC{}, which improves performance using the ECAPA-TDNN speaker encoder. The experimental results demonstrate both \resemblyzerVC{} and \ecapatdnnVC{} can achieve comparable performance to non-causal baseline systems on the VCTK dataset and two out-of-domain datasets. Furthermore, both proposed systems can be deployed on a single CPU core with 55 ms latency and 0.78 real-time factor. Our demo is available online\footnote{https://bohan7.github.io/ALO-VC-demo/}.

\end{abstract}

\section{Introduction}
\label{sec:intro}

The objective of voice conversion (VC) is to convert an utterance from a source speaker into a target speaker's voice while preserving the linguistic content. An any-to-any one-shot VC system has the potential to allow any source speaker to clone the voice of any target speaker using only one reference utterance. Various one-shot VC systems, including VQMIVC~\cite{wang21n_interspeech} and DiffVC~\cite{popov2021diffusion}, have demonstrated promising results in terms of speech quality and speaker similarity. However, when deploying the VC systems on actual hardware, it is critical to consider the low-latency implementation of the systems.

So far, numerous deep learning-based one-shot VC systems that can disentangle the source speaker's identity and content representation and clone the target speaker's voice characteristics have been developed. FragmentVC \cite{lin2021fragmentvc} employs Wav2Vec 2.0 to capture the latent phonetic structure of source speech, a convolution-based speaker encoder to extract speaker embedding from target speech, and the Transformer attention mechanism to predict the log mel-spectrogram of the converted speech. To reduce leakage of the source speaker's content information into the target speaker's representation, VQMIVC \cite{wang21n_interspeech} introduces vector quantization (VQ) for content representation and mutual information (MI) for optimization during training. Recently, diffusion-based one-shot VC system so-called DiffVC \cite{popov2021diffusion} applies a Transformer-based content encoder, speaker encoder, and diffusion-based decoder to achieve guiding results in speech quality and speaker similarity even on out-of-domain unseen speakers. The aforementioned non-causal systems contain a large number of parameters and use a huge number of future frames look-ahead. Hence, the latency exceeds the limited algorithmic latency of real-time applications (e.g., 60 ms). 

FastVC \cite{barbany2020fastvc} can perform many-to-many voice conversion four times faster than real-time on a single CPU core. However, it relies on future frames and cannot convert source speech in streaming. For practical real-time applications, FastS2S-VC \cite{kameoka2021fasts2s}, and Streamable VQMIVC \cite{yang2022streamable} are proposed to employ a streaming version of voice conversion. While FastS2S-VC can only convert the source speaker's voice into a limited set of target voices, Streamable VQMIVC can perform real-time streaming any-to-any voice conversion. However, Streamable VQMIVC adds two complex modules to the content encoder and pitch extractor of VQMIVC: (1) a content augmenter using two 2-D convolution layers and 12 Conformer blocks; (2) a pitch post-processing network consisting of 4 1-D convolution layers used to smooth the extracted streaming pitch. Therefore, Streamable VQMIVC contains many parameters (much more than VQMIVC) and a high level of complexity, making it difficult to deploy on current hardware. 

In this work, we propose the any-to-any low-latency one-shot voice conversion (\alovc{}) system based on the assumption that PPGs representation can help develop a high-quality causal system. To achieve better performance of low-latency voice conversion, we incorporate a pitch predictor, as well as causal positional encoding, to provide location information of a phoneme in PPGs. We introduce two versions of \alovc{} systems that can perform low-latency one-shot voice conversion for any-to-any speaker pairs using a look-ahead of 47.5 ms. \resemblyzerVC{} uses a pre-trained d-vector speaker encoder. We further improve the performance by using ECAPA-TDNN \cite{desplanques2020ecapa} speaker encoder, which results in the second version of \alovc{}, \ecapatdnnVC{}. These systems contain 8.3 million parameters, which is 3.2 times smaller than VQMIVC. The parameter count of the speaker encoder is not included since it is only used once before streaming inference. Our experimental results demonstrate that \resemblyzerVC{} and \ecapatdnnVC{} can achieve higher speech naturalness and speaker similarity than VQMIVC on VCTK \cite{veaux2016superseded} dataset and even on two out-of-domain unseen speech datasets. Moreover, they narrow the performance gap between the low-latency one-shot VC system and the diffusion-based one-shot VC system. To our knowledge, they are the first any-to-any low-latency one-shot VC systems capable of achieving comparable performance to non-causal systems on datasets other than VCTK \cite{veaux2016superseded}.

This paper is structured as follows. We summarise the state-of-the-art low-latency VC systems in Section~\ref{sec:related}. Then, we introduce the architecture of our proposed \alovc{} systems in Section~\ref{sec:methods}. The experimental results are demonstrated in Section~\ref{section:experiment}.

\section{Related work}
\label{sec:related}

\subsection{Existing low-latency VC systems}
Low-latency real-time non-parallel voice conversion (LLRT-VC) framework \cite{tobing2021low} can perform many-to-many voice conversion with 2 lookup frames. It can be executed faster than real-time with a $0.87–0.95$ real-time factor (RTF) when deployed on a single CPU core. This LLRT-VC framework is designed based on a cyclic variational autoencoder (CycleVAE) \cite{tobing19_interspeech, vcc20vaebaseline} spectral model and multiband WaveRNN with data-driven linear prediction (MWDLP) \cite{tobing21_interspeech}. In addition, the CycleVAE module is fine-tuned with waveform domain loss from the pre-trained MWDLP. The fine-tuning approach can help LLRT-VC achieve similar performance to non-causal CycleVAE with Parallel WaveGAN. However, for intra-lingual conversion, the LLRT-VC framework is outperformed by cascaded automatic
speech recognition (ASR) with TTS (ASR-TTS) \cite{Huang2020} and Nagoya University (NU) T23 system \cite{huang20b_vccbc}. Furthermore, when applying the LLRT-VC framework to new source or target speakers, fine-tuning is required.

Almost causal low latency VC (AC-VC) \cite{ronssin2021ac} can achieve comparable speech quality to ASR-TTS \cite{Huang2020} with only 57.5 ms future look-ahead and approximately 2 million parameters. AC-VC has been successfully deployed as a desktop application performing real-time any-to-many voice conversion using the JUCE framework\footnote{https://juce.com/}. However, the speaker similarity score of AC-VC is lower than that of ASR-TTS \cite{Huang2020} and CASIA \cite{zheng2020casia} systems due to the absence of future context. It is hypothesized that future frames are necessary to convert source speech prosody according to the target speaker’s prosody, and future information can help perform the consistent conversion \cite{ronssin2021ac}. AC-VC is an any-to-many voice conversion method and thus can only convert the voice of the source speaker to a limited number of target speakers. Therefore, to add a new target speaker, the entire system needs to be retrained.


In this work, we propose two versions of any-to-any low-latency one-shot VC system, which can perform voice conversion using a single utterance from any target speaker. Inspired by AC-VC, we also choose PPGs as the causal content representation. To convert the source speaker's prosody, a pitch predictor is used to predict the converted pitch trajectory, and pitch variance information is added to the PPGs representations. Moreover, we try to improve speaker similarity over AC-VC by using pre-trained speaker encoders from state-of-the-art speaker verification works. 


\subsection{LPCNet}
LPCNet \cite{valin2019lpcnet} is a state-of-the-art low-latency neural vocoder. It is composed of a frame rate network and a sample rate network that can synthesize highly quality waveform using only 30 ms future look-ahead. The frame rate network processes each 10 ms frame from the input 18 Bark-scale cepstral coefficients (BSCCs), and 2 pitch parameters (period, correlation). Then, the sample rate network operated at 16kHz generates speech waveform using the output conditioning vector from the frame network, the linear predicted sample, the previously generated excitation sample, and the previous output sample. Despite LPCNet being an auto-regressive model, LPCNet can be implemented in real-time on a single-core CPU and even mobile phones.

Our systems employ LPCNet to generate the converted speech. Therefore, the converted BSCCs, pitch and pitch correlation for every 10 ms are the output from the conversion model.






\section{Proposed methods}
\label{sec:methods}

\begin{figure*}[t]
  \centering
  \includegraphics[width=\linewidth]{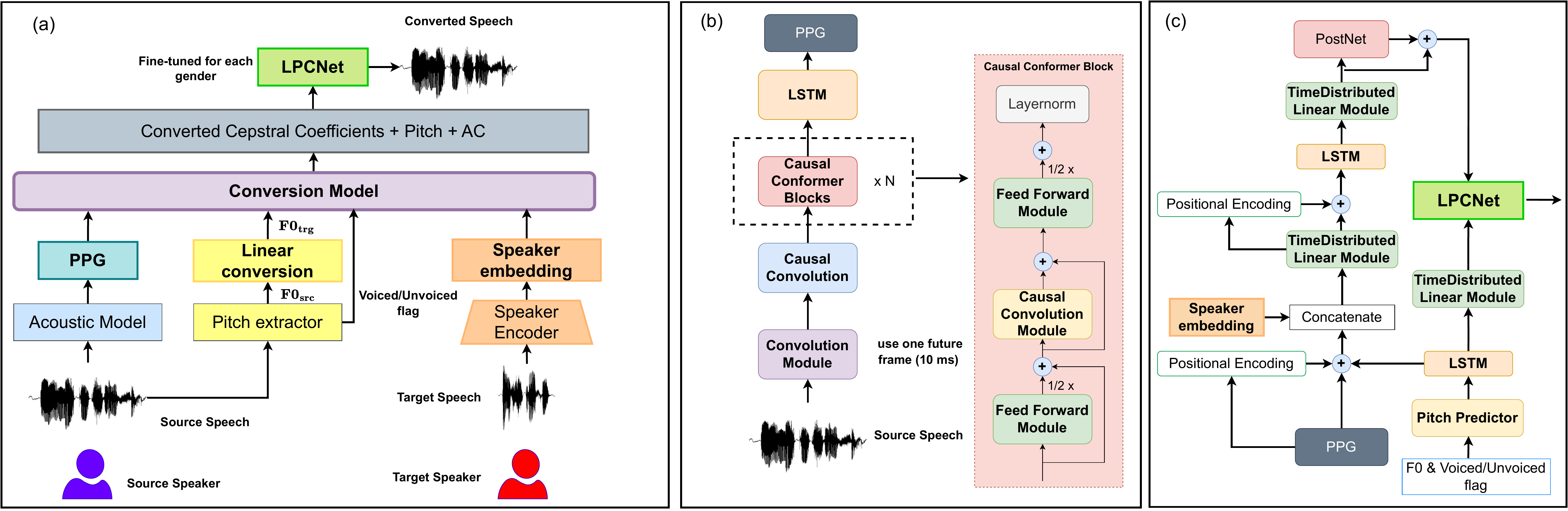}
  \caption{Schematic diagram of low latency PPGs based one-shot VC system. (a) the overview of the low latency one-shot VC system. (b) the schematic of Conformer based acoustic model (c) the schematic of the conversion model}
  \label{fig:schematic_vc_system}
\end{figure*}


Our proposed system, as indicated in Figure~\ref{fig:schematic_vc_system} (a), consists of five components, including an acoustic model, a pitch extractor based on speech signal processing, a speaker encoder, a conversion model, and LPCNet. In the following sections, we provide a detailed description of each component in our system. Additionally, we present a method to fine-tune the LPCNet model.

\subsection{Acoustic model}
The purpose of the acoustic model is to extract a speaker independent linguistic content representation of the source utterance. In this work, phonetic posteriorgrams are used. We design our acoustic model based on Conformer \cite{gulati20_interspeech} architecture, but without multi-head attention layers. Multi-head attention layers can increase the number of parameters and inference latency. As shown in Figure~\ref{fig:schematic_vc_system} (b), our acoustic model consists of a convolution module, a causal convolution module, causal conformer blocks, and uni-directional LSTM layers. The convolution module, which comprises a single 1-D convolution layer with a $3 \times 3$ kernel, detects the local correlation between adjacent frames. The module is followed by a causal convolution layer with ReLU activation, 4 causal Conformer blocks, two uni-directional
LSTMs and a time-distributed fully connected layer with softmax activation. To ensure causality in the causal conformer blocks, we perform layer normalization for each time frame independently. Finally, 512-D PPGs representation is extracted by the last LSTM layer. 

The acoustic model has a total of 2.7 million parameters and uses a 10 ms future look-ahead frame. The model is trained using cross-entropy loss between the ground-truth phoneme alignments \cite{lugosch19_interspeech, mcauliffe2017montreal} and the output from the acoustic model. Moreover, we trained the acoustic model separately from the conversion model to prevent the source speaker's identity from being exposed in the PPGs representation.



\subsection{Speaker encoder}
To obtain speaker embeddings, we use the pre-trained d-vector system \cite{wan2018generalized} from Resemblyzer\footnote{https://github.com/resemble-ai/Resemblyzer} and the pre-trained ECAPA-TDNN \cite{desplanques2020ecapa} model from SpeechBrain toolkit \cite{ravanelli2021speechbrain} respectively. The d-vector speaker encoder has been trained on VoxCeleb2 and contains 1.4 M parameters. The 14.7 M-parameter ECAPA-TDNN is trained on VoxCeleb1 \cite{Nagrani2017} and VoxCeleb2 \cite{chung18b_interspeech} datasets. The pre-trained models can reliably separate the identity and content representations of the target speaker. Furthermore, they are trained independently of the acoustic and conversion models, which prevents the leakage of content information. We propose two versions of \alovc{} system based on the two speaker encoders: \resemblyzerVC{} with a d-vector speaker encoder and \ecapatdnnVC{} with ECAPA-TDNN.



\subsection{Conversion model}
The conversion model shown in Figure~\ref{fig:schematic_vc_system} (c) is responsible for generating converted 18-dimensional Bark-scale cepstral coefficients (BSCCs), as well as converted pitch and voiced/unvoiced flag (VUF) for LPCNet to synthesize the converted speech.

The designed model includes a pitch predictor, two positional encoding modules, time-distributed fully connected layers, and uni-directional LSTM layers. The pitch predictor employs four 1-D causal convolution layers and one uni-directional LSTM, with converted pitch  $\text{F0}_{\text{trg}}$ and voiced/unvoiced flag (VUF) as inputs. The built-in pitch detector of LPCNet is used to extract F0 and VUF from the source speech. $\text{F0}_{\text{trg}}$ can be calculated with log-scaled $\text{F0}$ statistics of the target speech as follows:

\begin{equation}
\log \text{F0}_{\text{trg}} = \left(\log \text{F0}_{\text{src}} - \mu_{\text{src}}\right) *\left(\frac{\sigma_{\text{trg}}}{\sigma_\text{src}}\right)+\mu_{\text{trg}}
\end{equation}

where $\mu_{\text{src}}$ and $\mu_{\text{trg}}$ are the mean values of the source speaker's and the target speaker's $\log \text{F0}$. $\sigma_{\text{src}}$ and $\sigma_{\text{trg}}$ are the variance values of the source speaker's and the target speaker's $\log \text{F0}$. We convert the $\log \text{F0}_{\text{trg}}$ back to $\text{F0}_{\text{trg}}$ after the pitch conversion. 
Finally, one time-distributed fully connected layer projects the representations of $\text{F0}_{\text{trg}}$ and VUF into 2 dimensions. The pitch predictor is utilized to predict both the VUF and target speech prosody. The VUF and pitch variance information are also added to PPGs. 


The positional encoding is used to provide location information of a phoneme in PPGs. It can be implemented in streaming inference as follows:

\begin{equation}
\begin{aligned}
\mathrm{P}(\mathrm{L} + 1, 2 \mathrm{i}) & =\sin \left(\frac{\mathrm{L} + 1}{10000^{2 \mathrm{i} / \mathrm{d}}}\right) \\
\mathrm{P}(\mathrm{L} + 1, 2 \mathrm{i}+1) & =\cos \left(\frac{\mathrm{L} + 1}{10000^{2 \mathrm{i} / \mathrm{d}}}\right)
\end{aligned}
\end{equation}


where $\mathrm{L}$ is the length of previous input representation. $\mathrm{d}$ is the dimension of the output embedding space. $\mathrm{i}$ ($0 \leq \mathrm{i}<\mathrm{d} / 2$) is the index to map the position function for a dimension in the output embedding space. We can use $\mathrm{L}$ for the positional encoding in streaming inference, which ensures that no future information is used.

Finally, we use a PostNet consisting of three 1-D causal convolution layers with Tanh activation to post-process the output representation. During training, the conversion model is optimized using mean absolute error (MAE) between the ground-truth BSCCs and the output representation from the PostNet. Additionally, MAE between the ground-truth pitch and predicted pitch from the pitch predictor and MAE between ground-truth VUF and predicted VUF are used. The conversion model contains 5.6 M parameters.


\subsection{Fine-tuned LPCNet for each gender}
To further improve the VC system’s performance, the average LPCNet is fine-tuned for VCTK \cite{veaux2016superseded} male speakers and female speakers respectively. This involves training the base LPCNet with a $0.0001$ learning rate, using speech from only one gender. 


\section{Experiments}
\label{section:experiment}

\subsection{Experimental setup}

\textbf{Dataset} The experiments were conducted on LibriSpeech \cite{panayotov2015librispeech} and VCTK \cite{veaux2016superseded} datasets. The acoustic model was trained on the "train-clean-100" and "train-clean-360" subsets of LibriSpeech, which contained a total of 1172 speakers. 40 speakers (20 females and 20 males) from the "dev-clean" subset were not seen during training. The conversion model was trained on 104 speakers from the VCTK corpus. The test set includes six unseen speakers, consisting of three females and three males. 
\newline
\textbf{Model configurations} 
We resample each speech into 16kHz sampling rate, and randomly crop or pad a speech into a fixed length (5 seconds). Each audio is also chunked into overlapping frames of 25 ms with a hop size of 10 ms. Finally, for each frame, 13-D Mel Frequency Cepstral Coefficients (MFCCs) are extracted along with their delta values, giving a sequence of 39-D acoustic features.
\newline
\textbf{Training details}
We trained the acoustic and conversion models for 300 epochs with batch size 32, using the Transformer learning rate scheduler \cite{vaswani2017attention}. The Adam optimizer was used with an initial learning rate of 1e-7. The learning rate increases by 1e-7 per step for the initial 10,000 training steps and then decreases proportionally to the inverse square root of the step number.
\newline
\textbf{Baseline systems}
VQMIVC is the main baseline we use in our comparisons. VQMIVC has been shown to outperform other state-of-the-art any-to-any one-shot VC systems. We also report the performance of DiffVC, which has been shown to outperform the PPG-based VC systems on a diverse group of unseen speakers, as indicated in~\cite{popov2021diffusion}. Our primary aim is not to achieve better performance than DiffVC but rather to reduce the performance gap between low-latency VC systems and state-of-the-art non-causal VC systems.
\newline
\textbf{Metrics}
In Section \ref{subsubsection:subjective_test_VCTK}, we conducted subjective tests on the converted audio samples of VCTK \cite{veaux2016superseded}. The details of the subjective tests are in Section \ref{subsubsection:subjective_test_VCTK}. In Section \ref{subsubsection:objective_tests}, we employed widely-used deep learning models to evaluate speech quality and speaker similarity. Pre-trained MOSNet \cite{lo19_interspeech} and MOSRA \cite{elhajal22_interspeech} are used to predict the mean opinion score (MOS) for speech quality. For speaker similarity, we use pre-trained ECAPA-TDNN or d-vector speaker encoder to capture the representations from target speech and converted speech. Cosine similarity is computed for each pair of representations. The average and the standard deviation values of each experiment are reported in Section \ref{subsection:experimental_results}.

\subsection{Experimental results and analysis}
\label{subsection:experimental_results}

\subsubsection{Subjective test on VCTK}
\label{subsubsection:subjective_test_VCTK}
We designed the subjective tests following the Voice Conversion Challenge 2020 \cite{yi20_vccbc}. 10 participants 
undertook the subjective tests to rate the speech naturalness and speaker similarity. The evaluated systems are blind to each participant. The listeners initially rated the voice naturalness on a scale from 1 to 5 (Bad, Poor, Fair, Good, Excellent). Then, the participants listened to the converted audio sample and the audio sample from the target speaker and rated the speaker similarity of the pair on a 4-point scale: different speakers (absolutely sure), different speakers (not sure), same speaker (not sure), same speaker (absolutely sure). 2 unseen female speakers (p268 and p301) and 2 unseen male speakers (p252 and p256) are selected for the subjective tests. 30 audio samples converted by each VC system are evaluated. Each sample was assessed 8 times on average for speech naturalness and speaker similarity. The results of the subjective test are reported in Table \ref{tab:results_subjective_test}. The results indicate that \resemblyzerVC{} and \ecapatdnnVC{} can achieve better speech naturalness and speaker similarity than VQMIVC and reduce the performance gap between low-latency VC systems and DiffVC.

\begin{table}[th]
  \caption{MOS and speaker similarity of subjective test on VCTK.}
  \label{tab:results_subjective_test}
  \centering
\begin{tabular}{lcc}
\hline
\multicolumn{1}{c}{\multirow{2}{*}{Systems}} & \multicolumn{1}{c}{\multirow{2}{*}{MOS}} & \multicolumn{1}{c}{\multirow{2}{*}{Similarity}} \\
\multicolumn{1}{c}{}                         & \multicolumn{1}{c}{}                     & \multicolumn{1}{c}{}                            \\ \hline
Source speech    & $4.24 \pm 0.85$  & - \\
\rowcolor{baseline_color}
VQMIVC \cite{wang21n_interspeech}    & $2.36 \pm 1.01$ & $2.13 \pm 1.02$\\
\rowcolor{baseline_color}
DiffVC \cite{popov2021diffusion}    & $\bm{3.50 \pm 1.10}$  & $2.23 \pm 1.11$\\\hline
\resemblyzerVC{}    & $2.60 \pm 0.98$  & $2.18 \pm 1.00$\\
\ecapatdnnVC{}    & $3.04 \pm 0.92$  & $\bm{2.35 \pm 1.06}$\\\hline                       
\end{tabular}
\vspace{-4mm}
\end{table}


\subsubsection{Objective tests}
\label{subsubsection:objective_tests}
We perform the unseen-to-unseen experiment on the LibriSpeech dataset to evaluate our systems on a distribution that differs from that of VCTK. We consider 6 unseen speakers including 3 females (1988, 2277, and 2412) and 3 males (251, 652, and 777). There are 30 pairs and 150 converted audio samples. The performance for each system is reported in Table~\ref{tab:results_objective_test}.

\begin{table}[th]
  \caption{Objective tests of the unseen-to-unseen VC. ALO-VC systems outperform non-causal VQMIVC in quality and speaker similarity.
  }
  \label{tab:results_objective_test}
  \centering
\resizebox{\columnwidth}{!}{
\begin{tabular}{lcccc}
\hline
\multicolumn{1}{c}{\multirow{2}{*}{Systems}} & \multicolumn{2}{c}{Naturalness} & \multicolumn{2}{c}{Cosine similarity}\\
\multicolumn{1}{c}{}  & MOSNet \cite{lo19_interspeech} & MOSRA \cite{elhajal22_interspeech} & Speech brain \cite{ravanelli2021speechbrain}       & Resemblyzer              \\ \hline
\multicolumn{5}{l}{\textbf{LibriSpeech unseen speakers cross-dataset test}} \\
\rowcolor{baseline_color}
VQMIVC \cite{wang21n_interspeech} & $3.28 \pm 0.51$ & $3.52 \pm 0.62$ & $0.12 \pm 0.10$ & $0.60 \pm 0.00$\\ 
\rowcolor{baseline_color}
DiffVC \cite{popov2021diffusion} & $3.73 \pm 0.47$ & $\bm{4.79 \pm 0.28}$ & $\bm{0.28 \pm 0.17}$ & $\bm{0.76 \pm 0.00}$\\ \hline
\resemblyzerVC & $\bm{3.94 \pm 0.40}$ & $4.49 \pm 0.30$  & $0.18 \pm 0.14$ & $0.66 \pm 0.00$ \\
\ecapatdnnVC & $3.85 \pm 0.40$ & $4.59 \pm 0.17$ & $0.27 \pm 0.14$ & $0.64 \pm 0.10$ \\ \hline

\multicolumn{5}{l}{\textbf{Internal unseen speakers cross-dataset  speakers test}} \\
\rowcolor{baseline_color}
VQMIVC \cite{wang21n_interspeech} & $3.21 \pm 0.28$ & $2.78 \pm 0.70$ & $0.15 \pm 0.10$ & $0.64 \pm 0.00$\\
\rowcolor{baseline_color}
DiffVC \cite{popov2021diffusion} & $3.61 \pm 0.35$ & $\bm{4.58 \pm 0.26}$ & $\bm{0.30 \pm 0.10}$ & $\bm{0.84 \pm 0.00}$\\ \hline
\resemblyzerVC & $\bm{4.10 \pm 0.31}$ & $4.34 \pm 0.30$  & $0.18 \pm 0.10$ & $0.74 \pm 0.00$ \\
\ecapatdnnVC & $3.97 \pm 0.37$ & $4.45 \pm 0.24$ & $0.28 \pm 0.10$ & $0.73 \pm 0.00$ \\ \hline     

\end{tabular}
}
\vspace{-4mm}
\end{table}

We conduct one last experiment on an internal speech dataset to evaluate the proposed systems for real-world application. This experiment involved 6 unseen speakers (3 females and 3 males) who are not native English speakers. There are 30 pairs and 150 converted audio samples. Table~\ref{tab:results_objective_test} exhibits the performance of each system. 
Overall, the results demonstrate our proposed systems can outperform VQMIVC in speech naturalness and speaker similarity, and narrow the performance gap between low-latency VC systems and DiffVC.

\subsubsection{Real-time factors and latency}
We measure the runtime performance of our proposed systems on a single core of an Intel(R) Core(TM) i7-6850K CPU clocked at 3.60 GHz. The latency for each frame and real-time factor including input/output, PPG extraction, pitch conversion, speech conversion and waveform generation are reported in Table \ref{tab:latency_system}. All the results are for converting 5 seconds source speech. As shown in Table \ref{tab:latency_system}, our proposed systems could be used in real-time applications. 

\begin{table}[th]
\caption{Latency and real-time factor of the proposed systems.
}
\label{tab:latency_system}
\centering
\begin{tabular}{lcc}
\hline
Systems & Latency & Real-time factor \\ \hline
\rowcolor{baseline_color}
VQMIVC \cite{wang21n_interspeech}  &    Not causal     &        1.08          \\
\rowcolor{baseline_color}
DiffVC \cite{popov2021diffusion}  &    Not causal     &        63.15          \\ \hline
\resemblyzerVC  &    55.4 ms     &      0.78            \\ \hline
\ecapatdnnVC  &    55.3 ms     &        0.78          \\ \hline
\end{tabular}
\vspace{-4mm}
\end{table}



\section{Discussion}
In our one-shot VC experiments detailed in Section \ref{subsection:experimental_results}, the duration of the target speaker data varies from $1$ to $20$ seconds. Our experimental results in Section \ref{subsection:experimental_results} showcase the effectiveness of our pitch conversion method across varying target audio lengths. These results further indicate that \resemblyzerVC{} and \ecapatdnnVC{} can outperform VQMIVC on all three datasets. This is probably attributed to the pitch variance information incorporated by the pitch predictor, and the location information conveyed by the causal positional encoding. Moreover, the use of PPGs representation can help develop a causal system. Our work intends to encourage further research in real-time voice conversion tasks, which could facilitate the creation of voice avatars and recovery of impaired voice. Furthermore, the combination of PPGs representation and diffusion model for a causal system remains an unexplored area for future research.
\section{Conclusion}

This work proposes two versions of the any-to-any low-latency one-shot voice conversion (VC) system, so called \resemblyzerVC{} and \ecapatdnnVC{}, which use only 47.5 ms future lookahead. We also introduce a pitch predictor to predict the converted speech prosody, which helps the low-latency systems convert source speech pitch without future information. The pre-trained ECAPA-TDNN speaker encoders from state-of-the-art speaker verification works further improve speaker similarity. Our thorough evaluation with three datasets demonstrate \resemblyzerVC{} and \ecapatdnnVC{} can outperform VQMIVC (a non-causal VC system), and close the performance gap to DiffVC (a non-causal diffusion based VC system). The proposed systems can perform voice conversion with 55 ms latency, and faster than real-time on a single CPU core. 







\balance
\bibliographystyle{IEEEtran}
\bibliography{mybib}

\end{document}